\documentclass{ar-1col}
\usepackage{url}
\usepackage[numbers]{natbib}
\begin{document}

\markboth{Stern and Murugan}{Learning without neurons in physical systems}

\title{Learning without neurons in physical systems}

\author{Menachem Stern,$^1$ and Arvind Murugan$^2$
\affil{$1$Department of Physics and Astronomy, University of Pennsylvania, Philadelphia, PA 19104; email: nachis@sas.upenn.edu}
\affil{$^2$Department of Physics, University of Chicago,\\Chicago, IL 60637; email: amurugan@uchicago.edu}
}

\begin{abstract}
Learning is traditionally studied in biological or computational systems. The power of learning frameworks in solving hard inverse-problems provides an appealing case for the development of `physical learning' in which physical systems adopt desirable properties on their own without computational design. It was recently realized that large classes of physical systems can physically learn through \emph{local} learning rules, autonomously adapting their parameters in response to observed examples of use. We review recent work in the emerging field of physical learning, describing theoretical and experimental advances in areas ranging from molecular self-assembly to flow networks and mechanical materials. Physical learning machines provide multiple practical advantages over computer designed ones, in particular by not requiring an accurate model of the system, and their ability to autonomously adapt to changing needs over time. As theoretical constructs, physical learning machines afford a novel perspective on how physical constraints modify abstract learning theory.
\end{abstract}

%

\maketitle


\section{INTRODUCTION}

\subsection{Physical learning}

\begin{figure}[t]
\includegraphics[width=5in]{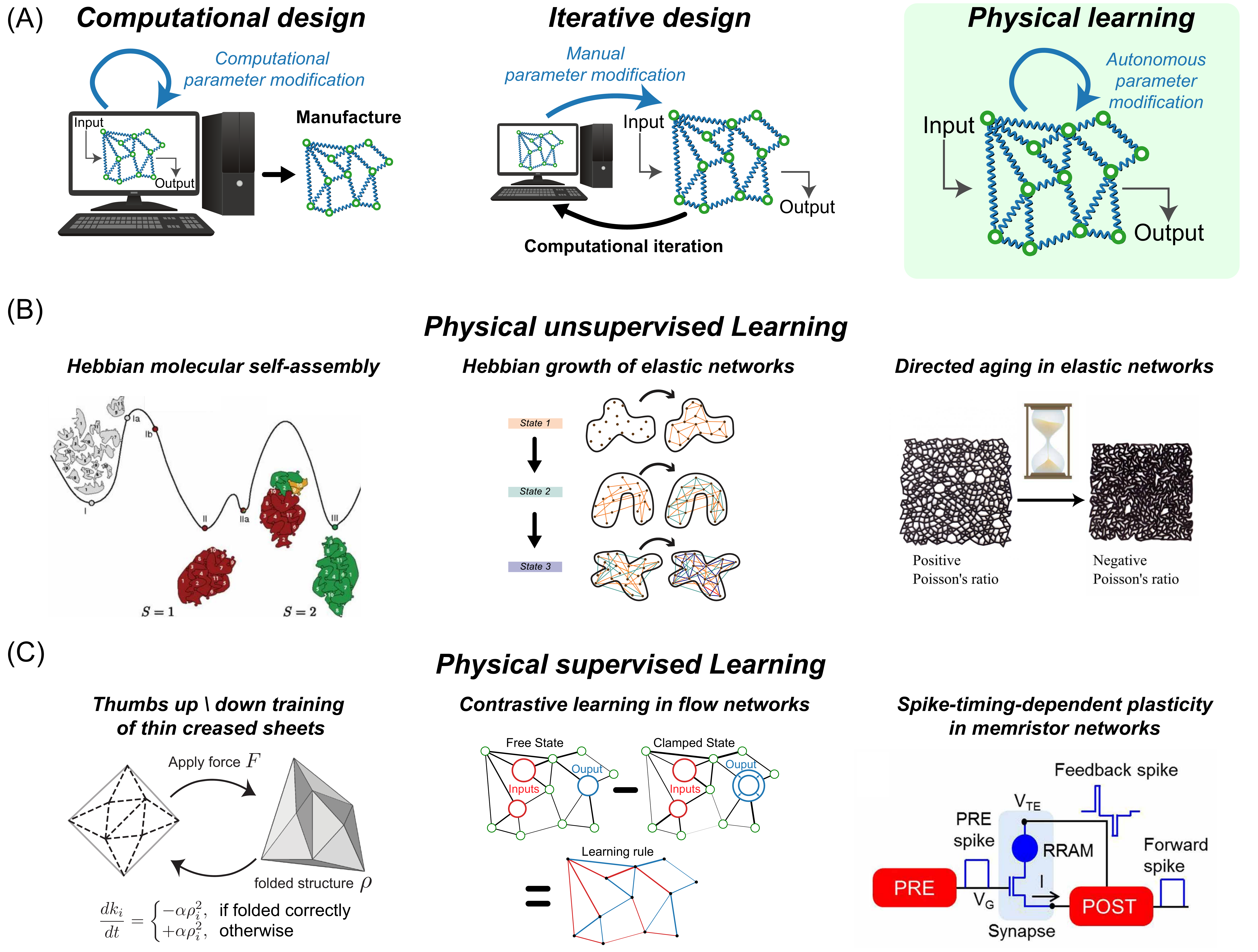}
\caption{Physical learning vs computer-aided design (A) Materials are often computationally designed for particular properties or responses, either entirely on a computer or through an iterative design-build-test process. In contrast, given external stimuli, in physical learning, materials autonomously modify their parameters to adopt desired properties or functions. Such autonomous learning machines modify themselves based on their response to stimuli according to physical `learning rules', which can be classified by the level of supervision.
(B) Physical unsupervised learning, e.g. Molecular self-assembly with Hebbian-learned interactions ~\cite{murugan2015multifarious}, Hebbian growth~\cite{stern2020continual} and directed aging in elastic networks~\cite{pashine2019directed} (credit to N. Pashine). (C) Physical supervised learning, e.g. Thumbs up/down rules in creased sheets~\cite{stern2020supervised}, contrastive learning in flow networks~\cite{stern2021supervised} and Spike timing dependent plasticity in memristive neural nets, reproduced from~\cite{pedretti2017memristive} (CC BY 4.0).}
\label{fig1}
\end{figure}

When no off-the-shelf solutions are readily available, systems are often rationally designed to provide solutions to our everyday problems. Designing such a system is known as solving an ``inverse problem'' since one attempts to find a system with a desired response to a perturbation
, in contrast to the ``forward problem'' of predicting responses of a given system to perturbations. 
To solve inverse problems, one must sift through possible systems by varying design parameters or design degrees of freedom (d.o.f). Solutions are typically sought using centralized, top-down approaches, where a designer can access and modify the elements of a system. Numerous computational algorithms have been developed for designing a wide range of physical systems ranging from materials to robots, including simulated annealing and genetic algorithms, that typically employ standard computers and may involve physical prototypes (Fig.~\ref{fig1}A). Inverse problems are usually much more difficult to solve compared to forward prediction problems, in particular because solutions are typically non-unique. 

\emph{Learning} is an established bottom-up framework for solving inverse problems~\cite{de2005learning} with specific advantages. Learning has been explored in the biological context of the brain and in artificial neural networks. However, learning is a more broadly applicable framework in which a system is changed incrementally in a way that enables adoption of a desired behavior. These incremental changes are defined as a function of the system’s response to external stimuli. A learning process consists of two major parts: (a) evaluating the system's output for a given input, (b) modification of the system based on the output. 
The system is first evaluated by applying stimuli from a training set. Then, the response of the system to the stimuli drives the modification of learning d.o.f, such that subsequent application of stimuli result in improved responses. 
This process is usually conducted iteratively until the system achieves satisfactory performance.

Here, we explore the emerging area of physical learning in materials. Typically, no computers are involved in this approach. Instead, a material is physically subject to examples of the desired behavior. In response, elements of the material 
- the \textit{learning degrees of freedom (d.o.f)} - change as described by a system-dependent dynamical process called a \textit{`learning rule’} (Fig.~\ref{fig1}A), underpinning the bottom-up, distributed nature of learning. A physical learning system thus has the following ingredients:
\begin{enumerate}
    \item Physical d.o.f $\mathbf{s}$ that respond to external stimuli $f$ by adopting a state or dynamic behavior $\mathbf{s}({f})$, a subset of which defines the desired behavior (i.e., output).
    \item Learning d.o.f ~$w_i$ that can modify how the physical d.o.f $\mathbf{s}(f;\{w_i\})$ respond to external stimuli $f$. 
    \item A \emph{learning rule}, $dw_i=h(\mathbf{s}(f;\{w_i\}))dt$, that modifies the learning d.o.f $w_i$ based on the response of the physical d.o.f $\mathbf{s}$ to the stimuli $f$. 
\end{enumerate}

Here, $h$ is an arbitrary non-linear function. The quality of learning is evaluated by whether the system learns the desired behavior, often quantified by a \emph{learning cost function} $C(\mathbf{s}(f;\{w_i\}))$. Note that the dynamics of the physical system itself do not compute the cost function; instead, physical learning seeks to minimize this cost function through the learning rule and the choice of stimuli used for training. 

Broadly, we can divide physical learning into categories based on the amount of user input during training: 
1) Physical unsupervised learning (Fig.~\ref{fig1}B), where a physical system self-adapts to instances of external stimuli with no supervisor intervention, 
typically by a Hebbian-like process~\cite{hebb2005organization}, and thereby learns to represent some aspect of these stimuli, 2) Physical supervised learning (Fig.~\ref{fig1}C), where an error signal is provided by a supervisor comparing the obtained and desired physical outcome, so systems can learn to exhibit specific responses for specific stimuli.


\subsection{Why learn using a physical system?}

Physical learning blurs the boundary between structure and function; the physical elements responsible for material properties also perform the task of adapting those material properties. Such a merger is motivated by both practical and theoretical considerations.

At a practical level, physical learning can lead to a new generation of materials capable of autonomous learning that offer several advantages over computer design. 1) While silicon-based computing and learning is incredibly powerful, a physical system that learns is more appropriate to perform tasks where the inputs and outputs are physical influences from or on the environment (e.g., forces, currents, molecule production) rather than symbolic information. 2) In this new framework, physical systems can learn from real examples of the desired behavior, so that the desired behaviors themselves do not need to be modeled on a computer. 3) Similarly, a detailed model of the physical system is not needed since the response of the real physical system, imperfections included, determines physical learning.
In contrast, computer-aided design is only as good as the model of the material and the model of the desired behavior and environment in which the material will be used. 4) Physical learning can allow systems to continually learn new behaviors \emph{in situ} as requirements change. 
5) Physical learning may be appropriate for applications with space, time or energy constraints or where robustness due to the distributed nature of learning and information processing (rather than in one electronic processor) is important.

From a theoretical perspective, studying learning in physical systems may shed new light on the fundamental requirements and limitations of learning in the presence of physical constraints. 
How can natural systems evolve the ability to learn? 
A theory of physical learning would expand our current modular conception of learning and memory in biological systems; we often look for control units (e.g., the brain) separable from the parts being controlled (e.g., the muscle). For example, even in single celled organisms, we tend to look for a separate gene regulatory or protein interaction circuits that learn and make decisions which are then carried out downstream by, say, physical self-assembly or cytoskeletal processes~\cite{micali2016bacterial}.  
But the downstream ``muscle'' itself can potentially learn and make decisions, as known to be the case in organisms ranging from ciliates to fruit flies and bee hives~\cite{bull2021ciliary, marbach2021network,ristroph2013active, peleg2018collective}. Physical learning could reveal the broader scope of such non-modular learning and information processing available for free in the ``muscle''~\cite{braitenberg1986vehicles}. 

Finally, this approach can provide a framework to understand \emph{atypical} disordered systems and how physical systems explore parameter space to arrive at such atypical points.
While we have made much progress in studying typical disordered systems through random ensembles, many examples of disordered systems in nature are highly atypical, e.g., due to evolution or other natural processes.  Physical learning provides one framework for such atypical phenomena as the origin of structures and architectures (e.g. hierarchies) in networks~\cite{kanai2015cerebral} and the prevalent low-dimensionality of physical responses~\cite{tlusty2017physical,husain2020physical}. 

The framework outlined here also builds upon ideas previously explored in materials science where materials are subject to some conditions (e.g., compression, temperature changes) with the goal of internal rearrangements that enhanced desired properties. For example, metals have been hardened by thermomechanical protocols~\cite{wang2002high} which reshape grain size and distribution in a way that improves strength, which in turn is part of a long tradition of using annealing and `hot stamping' protocols~\cite{karbasian2010review}. Similar ideas apply to processing polymers by which the same polymer mixes can be `trained' to rearrange molecules with different resultant mechanical properties by e.g. different extrusion protocols~\cite{baird2014polymer}. 

The framework here goes beyond such processing techniques by expanding the space of behaviors training can be used for, including supervised learning for input-output responses and pattern recognition of spatial or temporal correlations in chemical and mechanical stimuli. The framework also significantly expands the range of learning mechanisms beyond molecular rearrangements to include growth and degradation, and active learning of molecular interactions. Many of the advances here are a conceptual synthesis between ideas in materials processing and ideas of learning in computer science. A closely related development is the idea of memory in materials~\cite{keim2019memory}, discussed later. 

\subsection{Challenge and opportunity: local learning rules}



Machine learning algorithms implemented on a computer typically compute a cost function $C(\{w_i\})$, a global quantity that reflects performance on a task at hand, and strives to minimize it, e.g., through gradient descent,

\begin{equation}
\frac{dw_i}{dt} \sim - \nabla_{w_i} C(\{w_j\})  \quad {\rm (Non-local\ gradient\ descent)}.
\label{eq:gd}
\end{equation}

Gradient descent procedures are efficiently implemented on computers (especially GPUs), e.g. in artificial neural networks through back-propagation~\cite{lillicrap2020backpropagation}. Note that such a process is highly non-local when $w_i$ are interpreted as parts of a physical system: the change $dw_i$ in a given element $w_i$ depends in principle on the entire neural network, including e.g., how a `distant' neuron responded to stimuli. 

Such highly non-local changes cannot be realized in physical learning without effectively mimicking a computer algorithm with physical components. 
Instead, we seek to understand to what extent physical systems can learn by exploiting typically local natural processes without any explicit cost function,

\begin{equation}
\frac{dw[x,t]}{dt} \sim h(\mathbf{s}(f; \{w\})[x,t])  \quad {\rm (Local\ learning\ rule)}.
\label{eq:local}
\end{equation}

Here, the change in a learning degree of freedom $w[x,t]$ located at a point $x,t$ in the physical system is restricted to change only based on the state $\mathbf{s}[x,t]$ of the system at the same (or close vicinity) spacetime point (Here, $h$ is an arbitrary local function). 
Note however, that the local state $s(f;\{w\})[x,t]$ will generically depend on a stimulus $f$ and learning d.o.f $\{w\}$ of the entire system at distant points because of collective dynamics of the physical system; e.g., the way an elastic material deforms at point $x$ due to a force $f$ can depend on force components $f(y)$ and on material properties at distant points $y$.

Thus, there is no explicit cost function or global optimization involved in local rules in Eq.~\ref{eq:local}. However, if the local learning rule $h$ and stimuli $f$ are chosen correctly, the system can nevertheless minimize a cost function as the collective physical dynamics during the response to stimuli can encode global information in the local state $\mathbf{s}(f,\{w\})[x,t]$. 


Physical systems are in principle more constrained in their learning abilities than an \emph{in silico} neural network~\cite{nemenman2005fluctuation}. However, conceptually similar constraints also distinguish the brain from artificial neural networks; learning in the former is often constrained by locality (e.g., Hebb's rule~\cite{hebb2005organization} or Spike-Timing-Dependent Plasticity, STDP~\cite{caporale2008spike}). Nevertheless, biologically plausible learning rules, simulated on computers, have proven successful~\cite{bengio2015towards}, suggesting that physical learning has potential despite locality constraints~\cite{scellier2021deep}. Further, locality provides benefits as well, since learning can be desynchronous, more robust and scale better with system size since it does not rely on a central processor~\cite{dillavou2021demonstration,wycoff2022learning}.

\subsection{Relationship to machine learning, neuromorphic computing and physical computation}

There have been many threads of work related to physical learning over the previous years. 
The field of \textit{neuromorphic} computation~\cite{burr2017neuromorphic,markovic2020physics} is closely related to physical learning in that physical elements are modified so that the system adopts a desired computational ability. However, neuromorphic computers are often considered for a distinct objective - to compete with \textit{in silico} machine learning (ML) in its domain of symbolic inputs and outputs. The physical learning machines explored here can deal with problems where the inputs and outputs are physical (e.g., forces, molecular assemblies). Neuromorphic computing and \textit{in silico} ML algorithms do not directly compete in this realm since they require translation of stimuli to electronic or other symbolic inputs.  More saliently, these methods cannot directly display any physical output such as a elastic response or a self-assembled molecular structure in response to a stimulus.


The fields of \textit{physical computation}~\cite{piccinini2015physical} and \textit{molecular computation}~\cite{adleman1994molecular,soloveichik2008computation} explore how to implement fixed computations with physical systems. The systems explored here build on these prior works by autonomously learning what computation (or more generally, physical behavior) needs to be carried out in the first place by physically experiencing examples of the desired behavior. Another closely related physical computation framework is the field of \textit{reservoir computing}~\cite{tanaka2019recent}. Reservoir computing uses physical systems with such complex internal dynamics that no physical elements need to change during a learning process; instead, learning how to \emph{interface} with a fixed physical system can effectively define inputs and outputs in a way that solves inverse problems. This interface can be learned using a computer as an output filter~\cite{jaeger2004harnessing}.



In this review, we strive to showcase recent experimental and theoretical advances in the emerging field of physical learning. The remainder of the review is organized as follows: In section 2, we feature examples of physical learning machines, emphasizing ways of experimentally realizing physical learning. In section 3, we discuss the physically realizable learning rules powering learning machines, subdivided into unsupervised and supervised learning rules.  In section 4, we describe practical aspects inherent to realizing learning machines in experiments. In section 5 we highlight some theoretical learning concepts and how they manifest in the framework of physical learning. Finally, section 6 focuses on how the physical properties of a system are modified as it undergoes physical learning.

\section{EXAMPLES OF PHYSICAL LEARNING MACHINES}

\subsection{Elastic materials: networks and sheets}

Disordered elastic media and their abstractions, ranging from spring networks to creased sheets, have received much interest for their potential to mechanically respond in non-standard, desired ways~\cite{silverberg2014using,bertoldi2017flexible,rocks2017designing}. Recent works have explored the feasibility of physically learned behaviors in these systems~\cite{stern2020continual,pashine2019directed,hexner2019effect,hexner2019periodic,hagh2022transient}. Here, the learning d.o.f $w_i$ generally involve either bond lengths or bond stiffnesses in elastic networks and crease stiffnesses in creased sheets. 
For example, auxetic materials (negative Poisson's ratio, Fig.~\ref{fig1}B, right) were trained by imposing the desired global deformation (the stimulus $f$) on a material (Fig.~\ref{fig2}A), which leads to a pattern of local strain across the material (the response $\mathbf{s}(f)$). A learning process, typical of these works, involves bond stiffnesses $w_i$ that change according to the strain in that bond (i.e., locally) according to rules such as 
$\frac{dw_i}{dt}\sim - (\mbox{strain}_i)^2$. 

Physical learning can also train non-linear features like bifurcated folding pathways in mechanical systems (Fig.~\ref{fig1}C, left). Origami and Kirigami sheets with disordered crease patterns have rich folding topologies~\cite{dudte2016programming}, starting from $2d$ flat sheets with creases and slits, and ending with various folded shapes in response to given force patterns. While the crease geometry 
is usually fixed by fabrication, the bending stiffness of creases can serve as learning d.o.f $w_i$ and may change in response to folding strain. 
Theory and simulations have shown that folding pathways of such sheets can be trained for specific topologies with applications to dynamical control of folding ~\cite{stern2018shaping}, mitigation of misfolding pathways~\cite{stern2017complexity} and even classification of mechanical force patterns, analogous to neural networks~\cite{stern2020supervised} (Fig.~\ref{fig2}B). 



\subsection{Molecular systems and active matter}
Molecular systems with specific interactions show behaviors that define the complexity of life, ranging from computation and neural network-like information processing~\cite{adleman1994molecular, bray1995protein, winfree1998algorithmic, cherry2018scaling} to structural behaviors like self-assembly of complex and dynamic structures, targeted phase separation and active matter~\cite{zeravcic2017colloquium,biffi2013phase, needleman2017active}. While most work tends to assume fixed interactions, several works have explored ways of treating molecular interactions as learning d.o.f $w_i$ and thus learning complex molecular behaviors. 

\begin{figure}[t]
\includegraphics[width=5in]{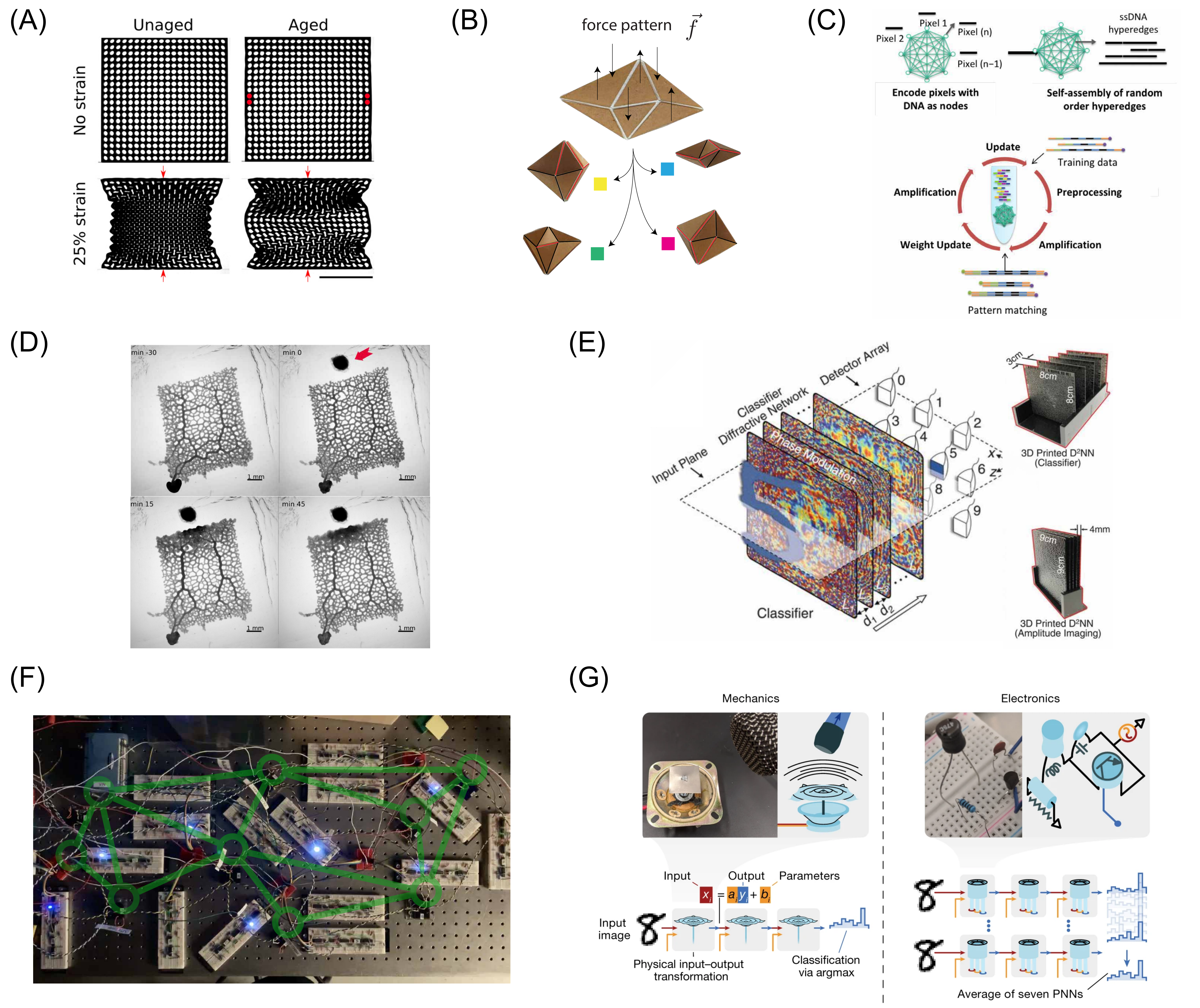}
\caption{Examples of experimentally realized physical learning machines. (A) Elastic network, aged under stress, adapts non-generic auxetic responses, reproduced from~\cite{pashine2019directed} (CC BY 4.0). (B) Self-folding origami sheets trained to fold into desired shape in response to an input force pattern. (C) DNA-based molecular learning for handwritten digit classification, reproduced from~\cite{baek2019enzymatic} (CC BY 4.0). 
(D) Pinned \textit{P. polycephalum} network changes its structure to memorize locations of food sources, reproduced from~\cite{kramar2021encoding} (Published under PNAS license).
(E) Optical diffractive neural network constructed to classify digits and fashion products, from~\cite{lin2018all}. Reprinted with permission from AAAS. (F) Electric circuit with variable resistors performs supervised learning and is robust to damage to individual elements~\cite{wycoff2022learning}. (G) Mechanical, electronic and optical networks trained to classify handwritten digits using Physics-aware-training, reproduced from~\cite{wright2021deep} (CC BY 4.0).  
}
\label{fig2}
\end{figure}

One broad learning mechanism involves creating new molecular species through ligation or polymerization~\cite{schaus2017dna,fredriksson2002protein} or by preferential multiplication of molecules that bind input molecules tightly and discarding of those who do not bind~\cite{lee2017vitro, baek2019enzymatic} (Fig.~\ref{fig2}C). These new species then mediate new interactions $w_i$ contingent on the history of the system~\cite{baum1995building,MILLS1999175}. One recent example~\cite{murugan2015multifarious,zhong2017associative} explores learning interactions needed for self-assembly using a Hebbian-inspired  `localized together, wired together’ rule (Fig.~\ref{fig1}B, left): $dw_{ij}/dt = s_i(x,t) s_j(x,t)$ where $s_i(x,t)$ is the concentration of species $i$ at $x,t$ and $w_{ij}$ is the interaction strength between species $i,j$. That is, species $i,j$ whose concentrations are high in the same place and time start interacting more strongly. The collective nucleation dynamics in such a system has been shown to be capable of pattern recognition by assembling different structures in response to different concentration patterns~\cite{zhong2017associative}, with mathematical connections to Hopfield's associative memory~\cite{hopfield1982neural}. 

Similar ideas of learning in multi-component phase separation ~\cite{jacobs2021self,shrinivas2021phase}, e,g., based on DNA nanostars~\cite{biffi2013phase}, are likely to be realized in the coming years. All of these works suggest that inevitable physical processes~\cite{winfree1998algorithmic} such as self-assembly, phase separation and nucleation can enable learning and perform pattern recognition, despite not being set up to mimic a neural network element by element~\cite{zhong2017associative} as in circuit approaches. 

Another important type of molecular learning involves growth. Molecular systems like hydrogels and DNA nanotubes can grow elements based on their current geometry~\cite{lee2012mechanical,Mohammed2017-il} and can potentially continually learn multi-stability~\cite{stern2020continual}. Some molecular systems such as crystals with defects are capable of rudimentary evolution through repeated fracture and growth; this aspect raises the intriguing possibility of learning through evolution in non-biological systems~\cite{schulman2012robust}.

Learning in well-mixed molecular circuits (as opposed to the structural processes above) has a larger literature, ranging from bio-inspired associative learning~\cite{gandhi2007associative} to recent progress towards molecular neural networks~\cite{lakin2016supervised,cherry2018scaling}. Molecular systems can also exploit their intrinsic stochasticity needed e.g., for Boltzmann machines~\cite{poole2017chemical}.


Active systems (e.g. self-propelled particles) can expend energy and perform actions on microscopic scales. Such systems are potentially capable of learning either by combining synthetic active matter with DNA-based elements~\cite{tayar2021active} and through rearrangement of nematic or polar filaments as in natural systems such as the cytoskeleton~\cite{saha2018joining,majumdar2018mechanical}. Related ideas have been explored through simulations~\cite{li2019particle,cichos2020machine,falk2021learning}, though truly autonomous learning has yet to be demonstrated on the molecular scale. Autonomous learning can be realized more readily in macroscopic robotic swarms~\cite{rubenstein2014programmable,werfel2014designing} that demonstrate similar active matter phenomena. Similarly, tissue-level rearrangements through active adhesion and junction changes might also allow for training~\cite{banerjee2020actin}.

\subsection{Flow networks}
In both biological (e.g. vascular) and engineered (e.g. microfluidic) networks, transport of materials is often achieved via fluid flow. Properties of pipes in a flow network, such as their radii, conductance or capacitance, determine the network's ability to globally transport material from one point to another. While such optimization could be carried out on a computer, many natural systems appear to adjust individual elements based on local feedback. For example, flow might couple to the mechanical properties of the pipes, constricting or expanding them to control flow conductance and pipe capacitance locally. Such adaptation might allow the network to control the distribution of cargo. For example, in \textit{Physarum polycephalum}, the thickness of tubes controls the shape of the organism and allows it to move, forage and memorize features of its surrounding~\cite{marbach2021network,kramar2021encoding,tero2010rules} (Fig.~\ref{fig2}D).

Similarly, adaptive processes in other natural flow networks such as those in leaves and vasculature~\cite{katifori2010damage, ronellenfitsch2016global} are thought to work through local rules, as there is no central controller.  Theoretical work on such flow networks has explored the ability to learn different behaviors through local rules (Fig.~\ref{fig1}C, right), including ML-like classification of stimuli~\cite{stern2021supervised,anisetti2022learning}.
\subsection{Neuromorphic computing}

A related but distinct thread of research is neuromorphic computation, which typically use solid state, electronic or optical elements to construct networks, inspired by neuronal or computational learning~\cite{burr2017neuromorphic,markovic2020physics}. Promising implementations include crossbar arrays~\cite{kim2012functional}, spintronic tunnel junctions~\cite{grollier2020neuromorphic} and phase change materials (PCMs) in photonic systems (Fig.~\ref{fig2}E).
In these systems, elements might adapt their resistance (e.g., memristors, potentiometers) as part of learning. 

While past neuromorphic systems tried to implement non-local gradient descent~\cite{rosenthfulal2016ly} or biological learning rules~\cite{serrano2013stdp} using complex elements, a new generation of neuromorphic implement simpler local learning rules based on contrastive learning~\cite{kendall2020training, martin2021eqspike}, allowing for regression and classification~\cite{dillavou2021demonstration,wycoff2022learning} (Fig.~\ref{fig2}F).
Neuromorphic computing seeks to compete in traditional machine learning (ML) domain on energy and robustness~\cite{sui2020review, rajendran2019low}. In contrast, the spirit of this review is to explore learning as a metaphor for how the response of atypical disordered physical systems to physical stimuli can change in functional ways; we do not seek to solve problems in the traditional ML domain with physical systems.

A recent new approach goes beyond engineered neural network-like architectures, taking the power of backpropagation to exploit the dynamics of complex physical systems.  In this approach ~\cite{wright2021deep}, physical parameters of a mechanical, electronic or optical system are updated by a backpropagation algorithm that observes the response of the system to physical stimuli. Such training effectively creates `deep physical networks', i.e., physical systems, capable of solving classification problems (Fig.~\ref{fig2}G). Computer-aided backpropagation in~\cite{wright2021deep} is not a focus of this review; we focus on training that is potentially implementable through local rules \emph{in situ} by physical dynamics, instead of computer-aided backpropagation. 

On the other hand, there are similarities. The goal of the molecular and mechanical systems reviewed here and the approach of \cite{wright2021deep} is to exploit the intrinsic dynamics of physical systems for learning and computation on physical stimuli, rather than force physical systems to mimic neural network architectures element by element.
\subsection{Reservoir computing: Training computational interfaces}

Physical computation approaches often use a physical system as a computational resource, and opt to only train an interface with that system. 
In reservoir computing, a complex dynamical network takes an external input and perform a generic high dimensional `computation'. A filter is then trained using a computer on the output of the system, such that a desired result is obtained~\cite{tanaka2019recent}. Different types of physical systems were used in this way, including electronic, fluidic, mechanical and biological systems, for a broad range of regression and classification problems.

\section{PHYSICAL LEARNING RULES}

In learning theory, learning problems are typically divided into conceptually distinct classes, unsupervised and supervised learning. A similar distinction is worth making in the materials context since the classes correspond to potentially different applications with different physical training protocols. 

\begin{figure}[t]
\includegraphics[width=4.9in]{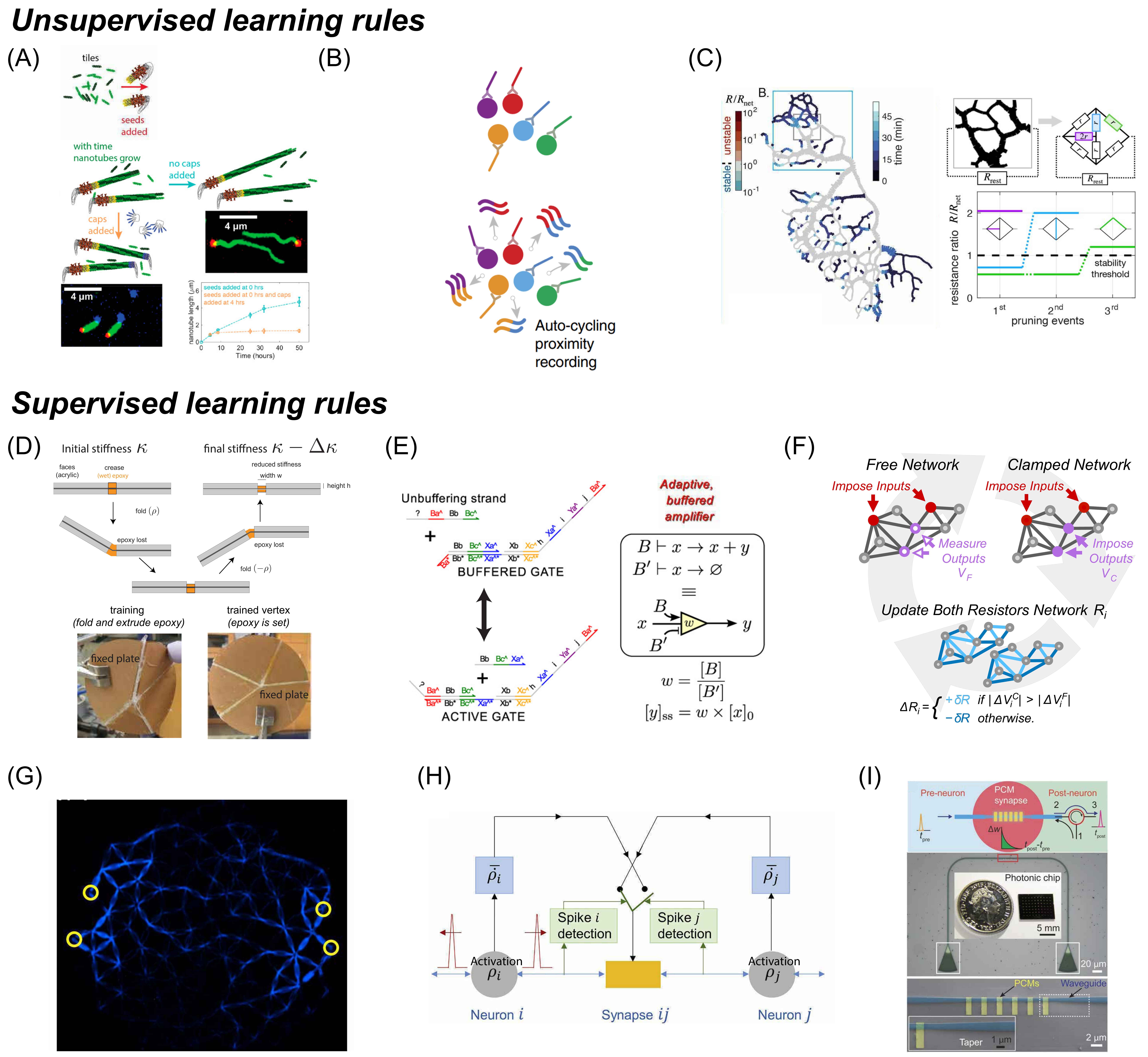}
\caption{Local unsupervised and supervised learning rules in various physical substrates. (A) DNA nanotubes self-assemble to reinforce connections between `seeds' based their spatial proximity. Adapted with permission from~\cite{agrawal2017terminating}. Copyright 2017 American Chemical Society. (B) Creation of interaction-mediating molecules based on spatio-temporal proximity of other molecular species; adapted from~\cite{schaus2017dna} (CC BY 4.0). (C) \textit{P. polycephalum} adapts the radius of vascular tubes according the local flow shear stress~\cite{marbach2021network} (credit K. Alim). (D) Epoxy-filled creases in sheets extrude epoxy in response to folding strain during the epoxy setting (i.e., training) period; consequently, final crease stiffnesses about the flat state depend on history of folding. (E) Buffered DNA strand displacement circuit feeds back into the concentration of reactants to implement gradient descent in supervised learning. Adapted with permission from~\cite{lakin2016supervised}. Copyright 2016 American Chemical Society.  (F) Two coupled electronic circuits, a free and a clamped network, that change the resistance of variable resistors in response to local voltage drops~\cite{dillavou2021demonstration}. (G) Local contrastive learning applied manually by pruning bonds in experimental photoelastic networks with visual stress measurements. Reprinted figure with permission from~\cite{pashine2021local}. Copyright (2021) by the
American Physical Society. (H) A spiking electronic circuit implementing the Equilibrium propagation learning rule to modify network synaptic weights according to temporal spiking patterns, reproduced from~\cite{martin2021eqspike} (CC BY 4.0). (I) Optical transmissivity of elements in an optical neural network changes according to STDP, by feeding back optical signals into waveguides with phase change materials, reproduced from~\cite{cheng2017chip} (CC BY 4.0).}
\label{figRules}
\end{figure}

\subsection{Unsupervised learning}
In unsupervised learning, the learning degrees of freedom are modified directly as a response to observed signals. A completely unsupervised physical learning rule is of the form,

\begin{equation}
\frac{dw_i}{dt} = h(\mathbf{s}(f;\{ w_j\}))  \quad {\rm (Unsupervised\ learning)},
\label{eq:unsup}
\end{equation}


where $\mathbf{s}(f;\{ w_j\})$ is the configuration the system adopts in response to an external stimulus $f$, assuming material parameters $\{ w_j\}$. $h$ is an arbitrary non-linear function. Here, learning d.o.f $w_i$ adjust themselves based on the response $\mathbf{s}(f; \{ w_j\})$. A paradigmatic example of such learning is directed aging in elastic systems. For example, elastic networks have been trained to be auxetic~\cite{pashine2019directed} by holding materials in the auxetic configuration as strained bonds soften according to a learning rule (Eq.~\ref{eq:unsup}). In this case, aging reduces the energy of a system held at its desired state $\frac{dw_i}{dt} \sim -\frac{\partial}{\partial w_i}E(\tilde{\mathbf{s}};w_i)$, where $E(\tilde{\mathbf{s}})$ is the energy of a desired state $\tilde{\mathbf{s}}$. Note that as the energy of elastic networks is a sum over individual bonds $i$, this is a local rule, where every bond $i$ changes according to the stress (energy) it carries. Similarly, at the molecular scale, stabilizing elements can grow between parts of a material that stabilize configurations the material is held in. For example, learning rules have been demonstrated experimentally in hydrogels using polymerases~\cite{lee2012mechanical} and DNA nanotubes~\cite{Mohammed2017-il,agrawal2017terminating} (Fig.~\ref{figRules}A) and theoretically modeled in~\cite{stern2020continual} to continually store multiple configurations in one mechanical system, without overriding previous ones, and retrieve configurations based on partial prompts.

Unsupervised learning can create new molecular interactions (Fig.~\ref{figRules}B). For example, proximity-based ligation~\cite{schaus2017dna,fredriksson2002protein} can create interaction mediating molecules for species $i,j$ by creating a ligated $i-j$ molecule; further, ligation is naturally localized in space and time by mass-action kinetics, resulting in a  Hebbian-like learning rule:
\begin{equation}
\frac{dw_{ij}}{dt} \sim s_i(x,t) s_j(x,t)  \quad {\rm (Molecular \ Hebbian\ learning)},
\label{eq:HebMol}
\end{equation}
where $w_{ij}$ is the interaction strength between molecular species $i$ and $j$ with concentrations $s_i(x,y),s_j(x,t)$. Thus, spatial and temporal proximity during learning leads to changes in stronger or weaker binding interaction between different species of molecules. Equivalent mechanisms include activation of inactive particles e.g., through strand displacement~\cite{schaus2017dna} or phosphorylation.  Such learning was shown to allow ~\cite{murugan2015multifarious,zhong2017associative,bisker2018nonequilibrium} learning multiple self-assembly behaviors and recognize patterns through nucleation (see Fig.\ref{figSup}D). 

Adaptation in flow networks have also been modeled~\cite{marbach2021network,ronellenfitsch2016global} with unsupervised rules that only depend on the flow through edges, proportional to a pressure drop $s_j-s_k$:
\begin{equation}
\frac{dw_{ij}}{dt} \sim h(s_i - s_j)  \quad {\rm (Flux-based\ rule)},
\label{eq:FlowRule}
\end{equation}
where $w_{ij}$ is the conductance of an edge connecting nodes $i$ and $j$. In \textit{P. polycephalum}, the stimulus $f$ might correspond to food sources. Edges that exhibit low effective conductance are pruned, as observed in simulations and experiments~\cite{marbach2021network} (Fig.~\ref{figRules}C).

\subsection{Supervised learning}

In machine learning (ML), supervised learning involves labeled training examples, e.g., labeled images of cats and dogs. 
In the context of physical learning, supervised learning corresponds to training a material to show desired specific responses for all input stimuli $f$. A typical supervised learning task is classification (e.g. the Iris dataset~\cite{Iris}), where a system is trained to produce a specific response $s_A(f)$ for all stimuli $f \in F_A$ in class $F_A$ and a different response $s_B(f)$ for stimuli $f \in F_B$ in class $F_B$. Here, $F_A$ and $F_B$ are user-specified sets of stimuli and the challenge is two-fold: the physical learning process must identify correlations or features that separate the stimuli in $F_A$ and $F_B$ and further, must result in a material that responds only to these features in a specified way (Fig.~\ref{figSup}A). 

As in ML, one can evaluate learning performance by a cost function $C(\{w_i\})$ that is a measure of the number of stimuli $f$ for which the system (with given $\{w_i\}$) evokes an incorrect response. Then, the goal of physical supervised learning is for the material to naturally adapt $w_i$ such that the cost function $C$ is minimized.

To achieve such a goal in physical systems, one needs a `supervisor' who makes the learning process contingent in some way on the response $\mathbf{s}(f)$ to stimuli $f$ being `right' or `wrong' (as defined by the cost function). 



The simplest form of such supervision is a `thumbs up/thumbs down' feedback in which the supervisor only decides the sign of the autonomous unsupervised rules described earlier:

\begin{equation}
\frac{dw_i}{dt} = 
  \bigg\{ \begin{array}{cl}
    - h(\mathbf{s}(f;\{ w_i\}))  & \quad \textrm{if } \mathbf{s}(f) \textrm{ is correct}  \\
    + h(\mathbf{s}(f;\{ w_i\}))  & \quad \textrm{otherwise}
  \end{array}
  \quad\quad \textrm{(Thumbs up/Thumbs down rule)}
  \label{eq:semsup}
\end{equation}


Such a rule was shown to be powerful enough to learn and classify subtle force correlations applied to disordered mechanical systems~\cite{stern2020supervised}. In this theoretical study, spatial force patterns belonging to one of two sets $F_A,F_B$ are applied to a disordered creased sheet; a supervisor decides whether the sheet is folded into a desired structure $s_A$ or $s_B$ for stimuli in $F_A$ or $F_B$ respectively. If the folded structure is correct for the stimuli, the supervisor lets creases soften according to their strain, say, by immersing the folded structure in one chemical environment. Otherwise, the supervisor immerses the folded structure in a different environment that \emph{stiffens} creases based on strain. A simplified version of this local rule was realized by letting a liquid glue flow out of creases that significantly fold in a desired configuration in creased sheets (Fig.~\ref{figRules}D).
%

Similar supervised rules can be implemented in molecular systems, e.g., by controlled strand displacement~\cite{lakin2016supervised} (Fig.~\ref{figRules}E). Through such mechanisms, e.g., the unsupervised molecular Hebbian learning rule in Eq.~\ref{eq:HebMol} can be modified so that  interactions between co-localized molecules are either strengthened or weakened depending on the judgement of a supervisor.

\subsubsection{Contrastive learning}
A more powerful supervised learning framework, requiring greater supervision but promising better results, is inspired by contrastive Hebbian learning~\cite{movellan1991contrastive}. In contrastive learning, a system is trained by observing the contrast between its current responses to an input $\mathbf{s}(f)$ and a nudged `improved' response $\hat{\mathbf{s}}(f)$. This nudging takes the form of additional (weak) forces or constraints, applied by the supervisor on the system, such that it better represents the desired configuration (lowering the cost function). For physical systems that are defined by an energy function $E(\mathbf{s};w_i)$, a simple contrastive learning rule takes the form

\begin{equation}
\frac{dw_i}{dt} \sim \frac{\partial}{\partial w_i}[E(\mathbf{s}(f)) - E (\hat{\mathbf{s}}(f))] \quad {\rm (Contrastive\ learning)}.
\label{eq:contrastive}
\end{equation}

\begin{figure}[t]
\includegraphics[width=5in]{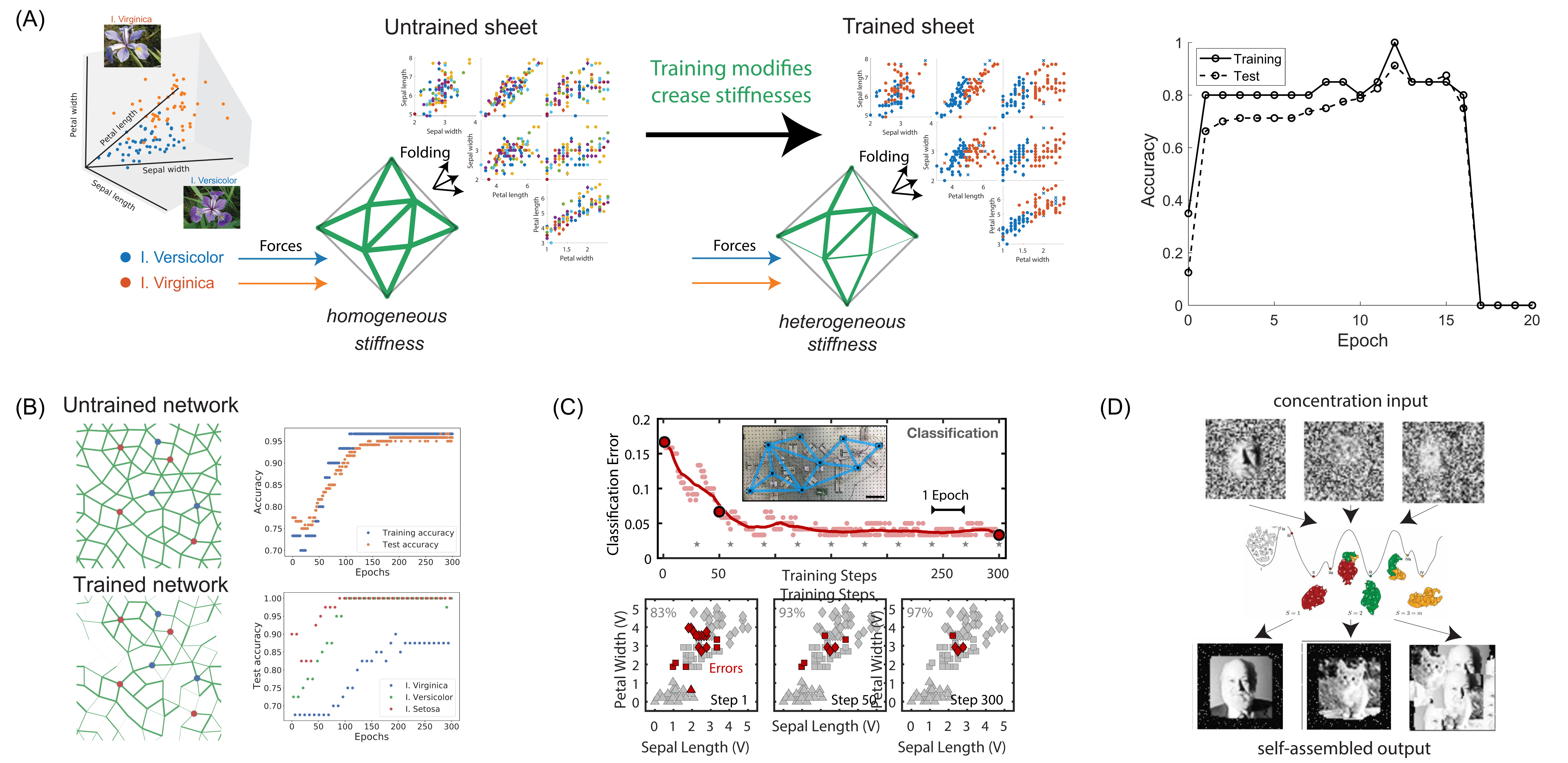}
\caption{Training physical learning machines for machine learning inspired tasks by mapping complex datasets to physical stimuli. (A) Data from the Iris flowers \cite{Iris} dataset is mapped to physical stimuli (here, force patterns) that are then used to train a creased thin sheet to show distinct responses for distinct classes of stimuli. Here, the sheet adopts a heterogeneous crease stiffness profile through a local learning rule, and eventually correctly classifies the input stimuli through folding responses~\cite{stern2020supervised}. (B,C) Similar classifications be learned in flow and electric networks by mapping the Iris data to input pressures or voltages at some nodes, and reading out the pressures or voltages at some other output nodes, in (B) simulations and (C) experiments~\cite{dillavou2021demonstration}. (D) 2500-pixel images were mapped to concentration patterns of 2500 molecular species (grayscale value of pixel $i$ = molecular concentration $a_i$). Molecular interactions were trained by a Hebbian rule, so the molecular system classifies different (potentially distorted) stimuli by showing different self-assembly behaviors in the non-equilibrium nucleation-dominated regime of self-assembly~\cite{zhong2017associative}.}
\label{figSup}
\end{figure}

While this approach is frequently used to train computational models (e.g. restricted Boltzmann machines~\cite{hinton2006fast}), it was shown to give rise to local, physically realizable learning rules~\cite{bengio2015stdp, bengio2015towards}. Furthermore, in certain limits contrastive learning well approximates the gradient of a global cost function~\cite{scellier2017equilibrium}, suggesting that physical local learning is not fundamentally limited in performance compared to global computational ML. Contrastive learning and its derivative approaches were shown to successfully train (in-silico) various physical models, including Hopfield nets~\cite{scellier2017equilibrium}, flow networks~\cite{kendall2020training,stern2021physical} (Fig.~\ref{figSup}B), mechanical spring networks~\cite{stern2021supervised}, cell assemblies and photonic neural networks~\cite{lopez2021self}. 
Early experimental systems implementing contrastive learning in linear resistor and elastic networks (Fig.~\ref{figRules}F,G) demonstrated the success of this approach~\cite{pashine2021local,dillavou2021demonstration,wycoff2022learning}, e.g. in classifying the Iris dataset (Fig.~\ref{figSup}C). These experiments further hint at the potential scalability of physical contrastive learning to large networks and complex sets of tasks.

A plausible generalization to systems without an energy function (i.e. non-symmetric dynamical systems) but still at steady state is given by
spike-timing-dependent-plasticity (STDP~\cite{caporale2008spike})  ~\cite{scellier2018generalization}: 
$\frac{dw_{ij}}{dt} \sim h(s_j) \dot{s}_i$ 
where $w_{ij}$ is the strength of a `synapse' (coupling) connecting `neurons' $j$ to $i$, and $h(\cdot)$ a non-linear function. This learning rule was shown computationally to enable learning in non-symmetric spiking recurrent neural networks (Fig.~\ref{figRules}H) by continuously evaluating the activations of `pre-synaptic' and `post-synaptic' nodes~\cite{martin2021eqspike}. Such rules have been implemented in physical memristive substrates~\cite{kim2015experimental,serb2016unsupervised};
e.g., optical transmissivity of chalcogenide phase change materials (PCM) changes as a function of the number of pulses applied to, affording an approximate implementation of STDP in optical synapses~\cite{cheng2017chip} (Fig.~\ref{figRules}I).


\section{CHALLENGES IN IMPLEMENTING PHYSICAL LEARNING}

\subsection{Physical implementation of supervision}
The principal difficulty in realizing physical learning machines is the requirement that the underlying physics allows for a useful local learning rule.

Many unsupervised rules explored here may be implemented by inevitable processes in natural systems. For example, directed aging in elastic systems~\cite{pashine2019directed,hexner2019effect,hexner2019periodic} exploits inevitable aging processes in many materials. Similarly, growth-based rules in molecular systems naturally follow local geometry~\cite{stern2020continual} while flow networks in many natural systems can expand or constrict according to local flow properties~\cite{marbach2021network}.  Proximity-based ligation~\cite{schaus2017dna} allows for molecular interactions to increase based on spatial or temporal proximity, e.g., enabling learning of self-assembly behaviors by simply co-localizing particles in the desired arrangement during a training period~\cite{murugan2015multifarious,bisker2018nonequilibrium}.

Supervised learning places greater demands on the material since it requires the same physical system to employ a learning rule with either sign, e.g., strengthening or weakening interactions, depending on context. For example, thumbs-up-thumbs-down supervised learning in creased sheets~\cite{stern2020supervised} required the same elastic material to be capable of stiffening (thumbs down) or softening (thumbs up) due to strain, based on a thumbs-up-or-down `signal' from the supervisor. Such sign switching might be achieved by the supervisor placing the system in different chemical environments (or at different temperatures~\cite{pashine2019directed}) based on whether the system shows the right or wrong response to the a stimulus.

Greater supervision like contrastive rules require more to realize their promise: in addition to the sign problem, contrastive rules require two copies of the system and the ability to change learning d.o.f $w_i$ based on the quantitative difference between the systems. While such duplication might be possible in some cases~\cite{dillavou2021demonstration}, a more broadly relevant implementation could apply the free and clamped conditions sequentially~\cite{scellier2017equilibrium}, if learning elements have a memory component. 
Another implementation might involve physical signals of multiple modalities to obtain the contrastive signal; e.g., in flow networks, tracer particles could be injected at the output and advected by the flow, carrying the error signal information to the learning d.o.f~\cite{anisetti2022learning}. Finally, approximations of idealized theoretical learning rules might suffice in real materials; e.g., correctly getting just the sign of change in $w_i$ based on change in energy might be sufficient for contrastive learning $w_i$~\cite{dillavou2021demonstration,bernstein2018signsgd}. 


\subsection{Locality in solid and liquid-like systems}

While physical learning rules need to be local, the nature of the locality requirement varies across systems. Solid-like systems, e.g., elastic networks and sheets and flow networks, typically have fixed neighborhood geometry. Each learning element $w_i$ has a fixed location in space and time $x,t$ and locality implies that $w_i$ can only change based on the state of the system $\mathbf{s}[x,t]$ in the vicinity. 

Liquid-like systems, like molecular systems, have no fixed neighbors in space, since components typically diffuse freely like in molecular systems or re-arrange in a more limited manner as in jammed packings~\cite{hagh2022transient}. While training might seem difficult in such liquid-like systems without fixed neighbors, there are two ways forward. In molecular systems with many species of components~\cite{jacobs2021self,shrinivas2021phase,murugan2015undesired}, the learning d.o.f $w_{ij}$ are interactions $J_{ij}$ between species $i,j$. The locality constraint dictates that $J_{ij}$ only change based on \emph{transient} spatial or temporal coincidences of molecules $i,j$ as in Eq.~\ref{eq:HebMol}, e.g., as considered in proximity-based ligation and in Hebbian learning of self-assembly~\cite{murugan2015multifarious, zhong2017associative}.  Note that the resulting learned interaction graph, encoded by $J_{ij}$, has no real-space interpretation and does not need to be embeddable in 3 dimensions.

A distinct challenge arises in liquid-like systems with only a few species or components, such as jammed sphere packings and actin or microtubule networks. Here, training often relies on the geometric arrangements, e.g., memories stored in cytoskeletal networks under shear~\cite{hagh2022transient, majumdar2018mechanical}. The further extension of learning protocols to such liquid-like systems without fixed neighbors is important to broaden the scope of physical learning to all scales. 



\subsection{Exploiting noise}
Compared to computer algorithms, physical systems experience higher noise that might alter how they learn in response to external stimuli~\cite{ventriglia2002stochastic}. Noise can affect both the physical response $\mathbf{s}(f)$ to stimuli $f$ and the consequent change in learning d.o.f $w_i$. 

Such inevitable fluctuations in physical systems can be a blessing since several learning tasks require randomness. For example, molecular systems with finite numbers of molecules can naturally act as `chemical Boltzmann machines' needed for the unsupervised learning of probability distributions~\cite{poole2017chemical}. Similarly, Ref.~\cite{zhong2017associative} shows how the intrinsically stochastic nature of nucleation of crystals can solve complex pattern recognition problems, much the way stochastic local search algorithms solve satisfiability problems on a computer. Noise improves the robustness of encoded memories in disordered jammed packings~\cite{keim2011generic}.  Noise during physical training might also be beneficial for generalization, as often found with stochastic gradient descent in \textit{in silico} ML~\cite{bottou2003stochastic}. The physical learning rules discussed earlier have been shown to be generally robust to noise, with an error floor~\cite{dillavou2021demonstration,stern2021physical}. Potential mitigation strategies include modulating the learning rate and the magnitude of stimuli during training (Fig.~\ref{fig5}A).

The distributed nature of physical learning also intrinsically makes it robust to other noise sources, like malfunctioning learning elements or damage. Similarly, unlike \textit{in silico} ML, physical learning is generally asynchronous since different learning d.o.f $w_i$ are updated independently without a central clocked processor, which might be beneficial for discrete learning d.o.f ~\cite{wycoff2022learning}.

\subsection{Material constraints: dynamic range of weights $w$, over-training}

A significant constraint on learning is the dynamic range of the learning d.o.f $w$ achievable in real materials without other compromises. For example, interactions in molecular systems can vary over a finite range (relative to $k_B T$) before associations become irreversible. Similarly, in mechanical systems, the range of bond stiffnesses might be limited, e.g., to avoid fracture. Such constraints might be overcome to some extent by choice of materials, such as shape memory polymers~\cite{mcknight2005variable,lagoudas2008shape}, hydrogels and poly-carbonates~\cite{henke2014high,zhou2019biasing}. Existing works~\cite{murugan2015multifarious,stern2020supervised, stern2021physical} have found that requirements for tasks studied so far are within experimentally available ranges. Another solution is through architecture - e.g., larger networks with more learning elements, each having a moderate dynamic range, might alleviate the need for larger dynamic ranges.

Some learning d.o.f $w_i$, such as stiffnesses, are constrained to be positive, an issue discussed before for the brain~\cite{sorscher2019unified}. Solutions proposed in ML, like shifting the task to positive values~\cite{hu2016dot}, or decomposing negative weights as the difference of two positive weights~\cite{wang2019reinforcement}, might have physical analogs as well. For example, positive molecular binding energies can be effectively shifted to negative values by adjusting entropic costs of binding~\cite{murugan2015multifarious}.

Another failure mode for learning physical systems is overtraining~\cite{stern2020supervised,amari1995statistical}. In materials that lose the variability of the learning d.o.f, overtraining can result in a network that can no longer adapt to new tasks~\cite{hexner2022loss}.

\section{PHYSICAL MANIFESTATIONS OF LEARNING CONCEPTS}

\subsection{Memory, learning and generalization}
A closely related concept to learning is memory. In the materials context, memory has a longer history of research, especially in disordered mechanical systems, polymer melts and glasses~\cite{keim2019memory}. Examples include colloidal systems `remembering’ the amplitude of prior shear and glasses `remembering’ a temperature at which they were aged. 

What is the relationship between learning and memory? On one hand, memory is a manifest requirement for learning; the learning d.o.f $w_i$ must encode a memory of past stimuli $f$. On the other hand, successful learning places additional demands on the nature of memory. Learning requires that the memory have a specific physical retrieval mechanism, namely in response to stimuli for which a functional response is sought. Further, learning often allows physical systems to \textit{generalize} and show the correct response to stimuli never seen before~\cite{murugan2015multifarious,stern2020continual,stern2020supervised, stern2021supervised,  cherry2018scaling}, much the way an \textit{in silico} neural network can generalize to novel cat images in a test set not seen during training. Such generalization requires selective memory for informative features in the stimuli; complete memory of all features of the training stimuli (over-fitting) would be counter-productive. In summary, learning requires selective and functional memory with a retrieval mechanism.

\begin{figure}[t]
\includegraphics[width=5in]{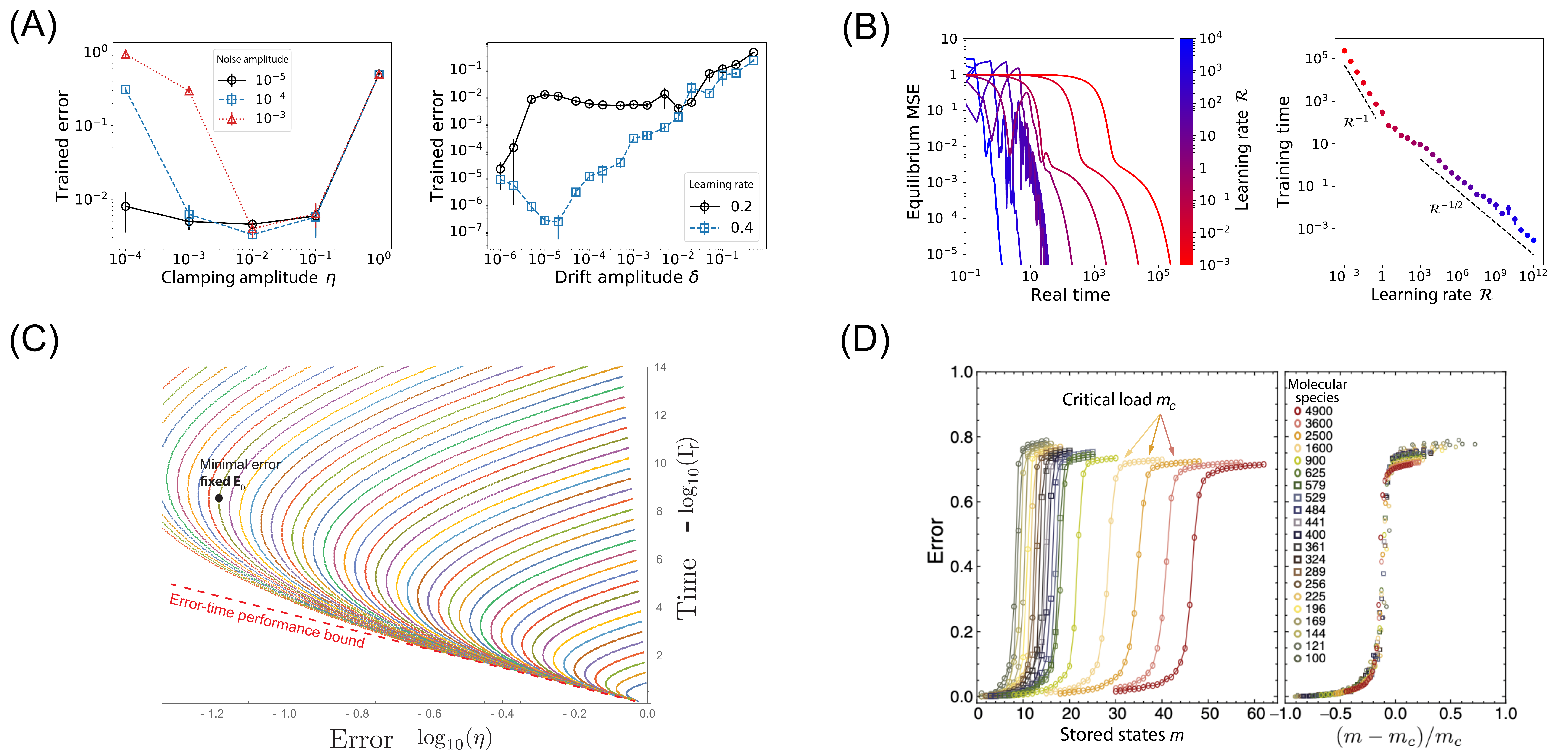}
\caption{Constraints on physical learning. (A) Noise in either the evaluation of the physical or learning d.o.f can impede learning, yet it can be mitigated by controlling physical hyper-parameters like the amplitude of clamping correction $\eta$, or the effective learning rate $\alpha$~\cite{stern2021supervised}. (B) Physical learning can succeed beyond physical equilibrium with minor effect as long as the learning rate does not exceed the physical dynamical rate $\mathcal{R}= 1$. Far beyond equilibrium, under-damped learning oscillations appear, yet the system might still be able to learn~\cite{stern2021physical}.
(C) Speed-accuracy trade-off in molecular systems trained to recognize patterns in molecular concentrations~\cite{zhong2017associative}; higher temperatures lead to higher accuracy and higher complexity of patterns recognized but come at a price of slower computation through nucleation and growth.  (D) Learning capacity scales as the square root of the number of distinct molecular species for 2-dim molecular self-assembly~\cite{murugan2015multifarious}.
}
\label{fig5}
\end{figure}

\subsection{Learning out of equilibrium}

Any learning process is associated with two distinct timescales: (1) a response timescale $\tau_{response}$ associated with the physical d.o.f responding to stimuli $f$, i.e., mapping inputs to outputs $f \to \mathbf{s}(f,\{w\})$, and (2) a learning timescale $\tau_{learn}$ associated with updating the learning d.o.f $w_i$ based on the response, i.e., $dw_i/dt \sim \frac{1}{\tau_{learn}} g(\mathbf{s}(f,\{w\}))$. \textit{In silico} ML has a clear separation between these timescales as the first process is typically fully completed before weights are updated.

In contrast, physical and biological systems might not have a clean separation between $\tau_{learn}$ and $\tau_{response}$; learning might effectively occur out-of-equilibrium and not be separable from the physical dynamics of stimuli-response. However, learning can be successful despite the lack of separation of timescales, as investigated for neuronal networks~\cite{zucker2002short,marom2010neural, ernoult2019updates,bartunov2018assessing} and tested recently for physical learning with contrastive rules~\cite{stern2021physical}. Even learning rates comparable to the physical response rate ($\tau_{learn} \sim \tau_{response}$) have little effect on the trained performance, though higher rates might lead to oscillations (Fig.~\ref{fig5}B). 






\subsection{Out of equilibrium effects: Time reversal symmetry, steady states, non-reciprocal interactions}

Beyond the relative rate of learning and physical dynamics discussed above, there are two distinct notions of being out-of-equilibrium relevant for learning: (a) the non-equilibrium nature of the learning process, (b) the non-equilibrium nature of the underlying physical system (in the absence of learning).


(a) The learning process: Successful learning is almost always out-of-equilibrium since it is irreversible; learning starts with a random set of learning d.o.f and ends with a distinct set of learning d.o.f $w_i$ from which it is likely impossible to reconstruct the initial $w_i$ exactly. As Landauer pointed out~\cite{landauer1961irreversibility}, such a many-to-one procedure in the space of weights $w_i$ can be interpreted as erasure and must consume free energy. Currently, it is unclear exactly how many-to-one this process is for different materials and tasks. 
At a more practical level, the learning process is bound to be dissipative in nature. For example, the molecular versions might involve polymerization, ligation or strand displacement~\cite{lakin2016supervised,schaus2017dna,baek2019enzymatic} while mechanical systems involve dissipative re-arrangements in foams and other complex materials~\cite{pashine2019directed}. The theoretical and practical constraints on learning by dissipation deserve further study. 

(b) The underlying physical system can be characterized as being at equilibrium or not without accounting for the learning process itself; many open questions remain on the distinction between the two cases.
For example, equilibrium systems with time-reversal symmetry can typically avail of more highly supervised protocols such as the contrastive Hebbian rule and echo backpropagation~\cite{movellan1991contrastive,lopez2021self}. Some of the systems studied here - e.g., elastic and flow networks - are often studied at steady state and preserve time-reversal symmetry. Consequently, influences can propagate from output to input as easily as input to output, enabling contrastive learning (Eq.~\ref{eq:contrastive}). In systems without time-reversal symmetry - e.g., molecular systems that are not at steady state or are powered by molecular motors, a perturbation of the output may not affect the input. Such systems can still be trained in an unsupervised way or through the thumbs-up-thumbs-down supervision described earlier. 

On the other hand, relaxing the constraint of being at equilibrium could potentially help learning. For example, the capacity of pattern recognition tasks in molecular systems may be improved using non-equilibrium nucleation-dominated self-assembly~\cite{zhong2017associative} and show associated trade-offs between complexity of pattern recognition, accuracy and speed of recognition (Fig.~\ref{fig5}C). Similarly, non-reciprocal interactions~\cite{fruchart2021non} should increase the range of learnable behaviors; for example, dynamic phases in Kuramoto-like networks of coupled oscillators could potentially be learned~\cite{seliger2002plasticity}. 
An open question is whether non-equilibrium behaviors are also more learnable; e.g., these systems may be more expressive~\cite{bisker2018nonequilibrium} or show greater degeneracy of parameters for a given behavior.

\subsection{Dynamic architectures, continual learning and forgetting}

Neural networks often have a fixed architecture (e.g., 2d convolutional network) with learning merely updating synaptic weights within that architecture. While some solid-state physical systems might be similarly constrained, physical systems that learn through growth or molecular interactions, have a freedom in architecture not usually present in \textit{in silico} ML. For example, growing networks of nanotubes or grown gels~\cite{Mohammed2017-il, agrawal2017terminating} or learned self-assembly~\cite{murugan2015multifarious} can change topology, geometry and even dimensionality during training. It is possible that learning can construct networks with architectures compatible to performing desired tasks. 

Further, physical systems naturally `forget' learned experiences, e.g., through degradation. Such erasure and forgetting can allow physical systems to naturally learn new tasks or functions without exceeding any capacity set by the number of d.o.f~\cite{stern2020continual,french1999catastrophic}.

\subsection{Expressivity, capacity, and hidden nodes}

A key property of any learning system is its `expressivity', i.e., the complexity of input-output relationships that can be modeled with the available learning d.o.f~\cite{bahri2020statistical}. A system with higher expressivity might also be easier to train. In fact, the success of ML is often attributed to overparameterization~\cite{rocks2022memorizing}. A related quantification is `capacity'~\cite{hopfield1982neural,hertz2018introduction}, e.g., related to the largest number of distinct behaviors that can be learned simultaneously.

As found for neural networks, both expressivity and capacity increase with the number of d.o.f for physical learning~\cite{murugan2015multifarious} (Fig.~\ref{fig5}D). Work in specific systems like learned self-assembly~\cite{murugan2015multifarious,zhong2017associative,fink2001many} shows that frustrated interactions and dimensionality of these d.o.f can dramatically increase expressivity and capacity. 
However, we do not yet have broad principles for controls expressivity are and what kinds of physical interactions increase it.


Another way to increase expressivity and capacity is the introduction of hidden nodes that do not directly couple to input stimuli or the output behavior, as in restricted Boltzmann machines~\cite{hinton2006fast}. In the materials context, similar ideas might apply, e.g., applying force patterns to only boundary and not the bulk~\cite{coulais2016combinatorial,pinson2016signal} or mapping inputs only to some molecules~\cite{cherry2018scaling,zhong2017associative}, might allow learning of more complex behaviors.

\section{PHYSICAL SIGNATURES OF PAST LEARNING}

Physical learning leads to disorder as the learning d.o.f $w_i$ typically become heterogeneous during learning. However, while the usual approach to disordered systems averages over random ensembles~\cite{mezard1987spin}, the learning process can lead to atypical instances of disorder. Here we discuss some signatures of learning in physical systems. 

We note that many of these learning signatures are analogous to signatures of evolution and evolvability~\cite{wagner1996perspective}. In both cases, functional systems are arrived at through an iterative procedure in parameter space $w_i$. Hence learning (and evolved) systems may be highly atypical in the context of `random' ensembles typically studied in statistical physics
; instead, these systems occupy functional regions accessible by the dynamical learning process.
The requirement of accessibility results in features that may be incidental or irrelevant to the task being trained or selected for; such features would not be present in random systems or systems designed by a highly non-local algorithm in parameter space. Much like with evolution, these signatures can be used to examine if a natural disordered system is the product of physical learning. 

\begin{figure}[t]
\includegraphics[width=5in]{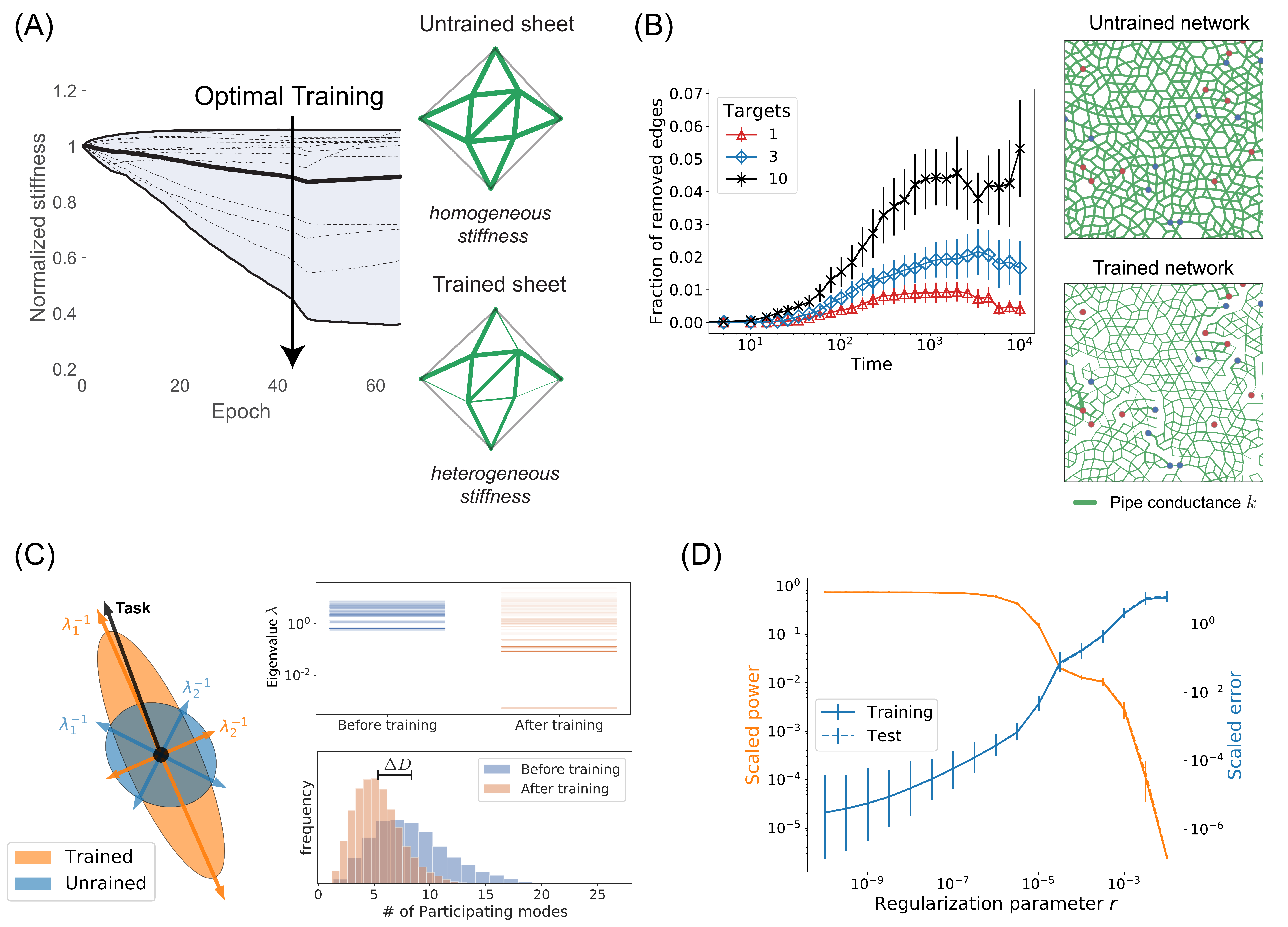}
\caption{Learning creates physical signatures in the substrate beyond the desired functionality. (A) Learning induces spatial heterogeneity in the physical substrate, here a self-folding sheet with stiff creases~\cite{stern2020supervised}. (B) System connectivity and topology may significantly change, as learning effectively prunes many unused or counterproductive edges in a network~\cite{stern2021supervised}. (C) Physical learning creates soft modes in a system, so that most forces couple preferentially to these few modes, reducing the effective response dimension $D$ of the system. (D) Depending on the difficulty of the task, a learning system may trade off performance for energy. Here, a regression task performed by a flow network shows such a trade-off when a parameter $r$ that regulates power consumption is varied.}
\label{fig6}
\end{figure}

\subsection{Network geometry and topology}

\textbf{Heterogeneities:} In solid-state systems such as elastic or flow networks, a learning process can leave profound signatures on the real space architecture. For example, like with synaptic weights in a neural network, learning inevitably leads to spatial heterogeneity in local elastic moduli or tube radii (Fig.~\ref{fig6}A), up to complete pruning of unused edges (Fig.~\ref{fig6}B). In liquid-state systems, the learning process can create atypical patterns of heterogeneous and promiscuous molecular interactions in chemical space; while a naive analysis might suggest a lack of function due to non-specificity~\cite{johnson2011nonspecific, huntley2016information}, the learned interactions are actually structured so as to be functional as seen in learned and evolved systems~\cite{murugan2015multifarious, su2022ligand}. Studying learning in relatively simple physical models may offer new insights on when and why such atypical heterogeneities develop, including potentially network motifs or hierarchical structures found in natural networks~\cite{kanai2015cerebral,alon2007network,modes2016extracting}.

\textbf{Adaptability:} Systems alternately trained for multiple incompatible behaviors in an alternating sequence can discover rare but highly adaptive or `mutable' networks for each task  \cite{kashtan2005spontaneous,hemery2015evolution,bloom2006protein,murugan2019bioinspired}; such adaptive networks are able to switch from one behavior to another with far fewer changes in parameters $w_i$ than would be typical for generic designed networks that perform those tasks.

\subsection{Network dynamics}
\textbf{Not-so-glassy landscapes:}
Generic frustrated many-body systems tend to have glassy landscapes with many randomly placed minima. However, learning can lead to exponentially fewer minima that are not randomly placed. For example, when molecular interactions are learned in a Hebbian way to self-assemble one of many structures, the number of resulting minima is exponentially fewer than naively expected~\cite{murugan2015multifarious}. Further, when the number of minima do proliferate near capacity, they are not random; rather, they correspond to chimeric assemblies of structures at other minima, analogous to spurious states in Hopfield associative memory. Similarly, jammed packings~\cite{hagh2022transient} can become ultra-stable by aging in that state while disordered creased sheets~\cite{stern2020supervised,stern2018shaping, stern2017complexity,pinson2017self}, upon training, show exponentially fewer branches at the flat state bifurcation point than expected of a random system. 

\textbf{Soft modes:}
Trained systems often show `soft' modes even if training did not explicitly seek such softness \cite{tlusty2017physical,husain2020physical,yan2018principles,recanatesi2021predictive}. Soft modes correspond to normal modes with low energy eigenvalue in equilibrium systems, or more generally, a small Lyapunov exponent. Consequently, trained systems respond more strongly to random forces than a random system; responses to random forces are often low dimensional along these soft modes~\cite{kaneko2003chaotic, transtrum2015perspective} (Fig.~\ref{fig6}C). Furthermore, the energy required in flow and resistor networks to actuate desired behaviors is often reduced by supervised learning
(Fig.~\ref{fig6}D).


\section{CONCLUSION}

In this review we discussed the general notions of autonomous physical learning machines, the physical constraints that affect them, and how some of them can be overcome. While clearly inspired by neuroscience and machine learning, we attempted to convey that physical learning is separate and unique: learning machines are physical, like biological learning networks, but can learn to solve inverse problems and produce responses that have no analogy in biology or machine learning. Moreover, treating learning as a physical phenomenon encourages research on fundamental questions on physically realizable learning models, how learning manifests as a collective behavior in systems with simple constituents, and how learning induces physical changes in these systems. We believe that such questions underlie a fundamental physical theory of learning, and that their answers may be independent of the specific details of implementation of the learning machine. Physical learning, both theoretical and especially experimental, requires multidisciplinary approaches. We hope recent and upcoming research in this field will interest not only condensed matter physicists, but computer scientists and neuroscientists as well.

\section*{DISCLOSURE STATEMENT}
The authors are not aware of any affiliations, memberships, funding, or financial holdings that
might be perceived as affecting the objectivity of this review. 

\section*{ACKNOWLEDGMENTS}
We thank Sam Dillavou, Constantine Evans, Martin Falk, Nidhi Pashine, Lulu Qian, Daniel Hexner, Shahar Yosha, Benjamin Scellier, Jason Rocks, Sid Nagel, Andrea Liu, Douglas Durian, Heinrich Jaeger and Erik Winfree for insightful discussions and suggestions.
M. S. would like to acknowledge funding from the National Science Foundation via grant DMR-2005749 and from the National Institute of Health CRCNS via grant 1R01MH125544-01.
This work was primarily supported by the University of Chicago Materials Research Science and Engineering Center, which is funded by National Science Foundation under award number DMR-2011854.


\end{document}